\documentclass[sigconf,screen]{acmart}

\AtBeginDocument{%
  \providecommand\BibTeX{{%
    \normalfont B\kern-0.5em{\scshape i\kern-0.25em b}\kern-0.8em\TeX}}}
\usepackage{enumitem}
\usepackage{multicol}
\usepackage{colortbl} 
\usepackage{amsmath,amsfonts}
\usepackage{algorithm}
\usepackage{algpseudocode}
\usepackage{flushend}
\usepackage{ulem}
\usepackage[switch]{lineno}
            
\usepackage{hhline}
\usepackage{listings}
\definecolor{verylightgray}{rgb}{.97,.97,.97}
\lstdefinelanguage{Solidity}{
  keywords=[1]{anonymous, assembly, assert, balance, break, call, callcode, case, catch, class, constant, continue, constructor, contract, debugger, default, delegatecall, delete, do, else, emit, event, experimental, export, external, false, finally, for, function, gas, if, implements, import, in, indexed, instanceof, interface, internal, is, length, library, log0, log1, log2, log3, log4, memory, modifier, new, payable, pragma, private, protected, public, pure, push, require, return, returns, revert, selfdestruct, send, solidity, storage, struct, suicide, super, switch, then, this, throw, transfer, true, try, typeof, using, value, view, while, with, addmod, ecrecover, keccak256, mulmod, ripemd160, sha256, sha3}, 
  keywordstyle=[1]\color{blue}\bfseries,
  keywords=[2]{address, bool, byte, bytes, bytes1, bytes2, bytes3, bytes4, bytes5, bytes6, bytes7, bytes8, bytes9, bytes10, bytes11, bytes12, bytes13, bytes14, bytes15, bytes16, bytes17, bytes18, bytes19, bytes20, bytes21, bytes22, bytes23, bytes24, bytes25, bytes26, bytes27, bytes28, bytes29, bytes30, bytes31, bytes32, enum, int, int8, int16, int24, int32, int40, int48, int56, int64, int72, int80, int88, int96, int104, int112, int120, int128, int136, int144, int152, int160, int168, int176, int184, int192, int200, int208, int216, int224, int232, int240, int248, int256, mapping, string, uint, uint8, uint16, uint24, uint32, uint40, uint48, uint56, uint64, uint72, uint80, uint88, uint96, uint104, uint112, uint120, uint128, uint136, uint144, uint152, uint160, uint168, uint176, uint184, uint192, uint200, uint208, uint216, uint224, uint232, uint240, uint248, uint256, var, void, ether, finney, szabo, wei, days, hours, minutes, seconds, weeks, years},  
  keywordstyle=[2]\color{teal}\bfseries,
  keywords=[3]{block, blockhash, coinbase, difficulty, gaslimit, number, timestamp, msg, data, gas, sender, sig, value, now, tx, gasprice, origin},  
  keywordstyle=[3]\color{violet}\bfseries,
  identifierstyle=\color{black},
  sensitive=false,
  comment=[l]{//},
  morecomment=[s]{/*}{*/},
  commentstyle=\color{gray}\ttfamily,
  stringstyle=\color{red}\ttfamily,
  morestring=[b]',
  morestring=[b]"
}
\usepackage{pifont}
\usepackage{diagbox}
\usepackage{subfigure}
\usepackage{tcolorbox}
\usepackage{color}
\usepackage{microtype}
\usepackage{balance}
\usepackage{oplotsymbl}
\usepackage{subfigure}
\usepackage{siunitx}
\usepackage{array,framed}
 \usepackage{
   color,
   float,
   epsfig,
   wrapfig,
   graphics,
   graphicx,
 }
\usepackage{setspace}
\usepackage{amsfonts}
\usepackage{latexsym,fancyhdr,url}
\usepackage{enumerate}
\usepackage{algpseudocode}
\usepackage{graphics}
\usepackage{xparse} 
\usepackage{xspace}
\usepackage{multirow}
\usepackage{appendix}
\usepackage{microtype}
\usepackage{soul}
\soulregister\cite7 
\soulregister\ref7 

\usepackage{color}
\usepackage[T1]{fontenc}
\usepackage{fancyhdr}
\newcommand{\boxmargin}{1mm}
\newtcolorbox{myboxc}{
    colback=gray!15!white,
    arc = 0pt, outer arc = 0pt,
    boxsep=0pt, left = 3pt, right = 0pt, top = 0pt, bottom = 0pt, 
    leftrule=3pt, bottomrule=0pt,toprule=0pt, rightrule=0pt,
    left = \boxmargin, right = \boxmargin, top = \boxmargin, bottom = \boxmargin
}
\usepackage{subfigure}
\setcopyright{acmlicensed}
\acmDOI{10.1145/3650212.3680353}
\acmYear{2024}
\copyrightyear{2024}
\acmISBN{979-8-4007-0612-7/24/09}
\acmConference[ISSTA '24]{Proceedings of the 33rd ACM SIGSOFT International Symposium on Software Testing and Analysis}{September 16--20, 2024}{Vienna, Austria}
\acmBooktitle{Proceedings of the 33rd ACM SIGSOFT International Symposium on Software Testing and Analysis (ISSTA '24), September 16--20, 2024, Vienna, Austria}
\acmSubmissionID{issta24main-p788-p}
\received{2024-04-12}
\received[accepted]{2024-07-03}
\begin{document}

\title{Identifying Smart Contract Security Issues in Code Snippets from \textit{Stack Overflow}}


\author{Jiachi Chen}
\orcid{0000-0002-0192-9992}
\affiliation{%
  \institution{Sun Yat-sen University}
  \city{Zhuhai}
  \country{China}
}
\email{chenjch86@mail.sysu.edu.cn}

\author{Chong Chen}
\orcid{0009-0003-8423-6757}
\affiliation{%
  \institution{Sun Yat-sen University}
  \city{Zhuhai}
  \country{China}
}
\email{chench578@mail2.sysu.edu.cn}

\author{Jiang Hu}
\orcid{0009-0005-8842-7811}
\affiliation{%
  \institution{Sun Yat-sen University}
  \city{Zhuhai}
  \country{China}
}
\email{hujiang5@mail2.sysu.edu.cn}

\author{John Grundy}
\orcid{0000-0003-4928-7076}
\affiliation{%
  \institution{Monash University}
  \city{Melbourne}
  \country{Australia}
}
\email{john.grundy@monash.edu}

\author{Yanlin Wang}
\authornote{corresponding author}
\orcid{0000-0001-7761-7269}
\affiliation{%
  \institution{Sun Yat-sen University}
  \city{Zhuhai}
  \country{China}
}
\email{wangylin36@mail.sysu.edu.cn}

\author{Ting Chen}
\orcid{0000-0001-9165-8331}
\affiliation{%
  \institution{University of Electronic Science and Technology of China}
  \city{Chengdu}
  \country{China}
}
\email{brokendragon@uestc.edu.cn}

\author{Zibin Zheng}
\orcid{0000-0002-7878-4330}
\affiliation{%
  \institution{Sun Yat-sen University}
  \city{Zhuhai}
  \country{China}
}
\email{zhzibin@mail.sysu.edu.cn}

\begin{abstract}

Smart contract developers frequently seek solutions to developmental challenges on Q\&A platforms such as \textit{Stack Overflow} \textit{(SO)}. Although community responses often provide viable solutions, the embedded code snippets can also contain hidden vulnerabilities. Integrating such code directly into smart contracts may make them susceptible to malicious attacks. We conducted an online survey and received 74 responses from smart contract developers. The results of this survey indicate that the majority (86.4\%) of participants do not sufficiently consider security when reusing \textit{SO} code snippets. Despite the existence of various tools designed to detect vulnerabilities in smart contracts, these tools are typically developed for analyzing fully-completed smart contracts and thus are ineffective for analyzing typical code snippets as found on \textit{SO}. We introduce \textit{SOChecker}, the first tool designed to identify potential vulnerabilities in incomplete \textit{SO} smart contract code snippets. \textit{SOChecker} first leverages a fine-tuned \textit{Llama2} model for code completion, followed by the application of symbolic execution methods for vulnerability detection. Our experimental results, derived from a dataset comprising 897 code snippets collected from smart contract-related \textit{SO} posts, demonstrate that \textit{SOChecker} achieves an F1 score of 68.2\%, greatly surpassing GPT-3.5 and GPT-4 (20.9\% and 33.2\% F1 Scores respectively). Our findings underscore the need to improve the security of code snippets from Q\&A websites. 

\end{abstract}






\maketitle
\section{Introduction}
\label{sec:intro}
In recent years, smart contracts have catalyzed the development of many new applications, such as Non-fungible Tokens (NFTs)~\cite{yang2023definition} and Decentralised Finance (DeFi)~\cite{su2023defiwarder}. Due to the rapidly evolving of blockchain technology and the limited availability of online resources, developers often turn to Q\&A platforms such as \textit{Stack Overflow} \textit{(SO)}~\cite{stackoverflow} for development guidance. 
Such Q\&A platforms may facilitate knowledge exchange and may help to address the questioner's issue. However, the shared code snippets in answers can also embed hidden vulnerabilities, posing significant security risks when incorporated naively into smart contracts, especially by inexperienced smart contract developers.





Various methods, such as static analysis~\cite{oyente,smartcheck,slither,securify}, dynamic analysis~\cite{confuzzius} and formal verification~\cite{so2020verismart}, have been proposed to detect vulnerabilities in smart contracts, these approaches usually require a fully complete and compilable smart contract code. However, conducting such security analyzes directly on incomplete shared smart contract code snippets from \textit{SO} posts presents significant unsolved challenges. Consequently, when analyzing code snippets from \textit{SO} posts, these tools may fail for the majority of cases.  Recent studies have demonstrated the promising capabilities of Large Language Models (LLMs)~\cite{wei2022emergent} in code-related tasks, including code completion~\cite{eghbali2024hallucinator} and code generation~\cite{zhang2023planning}. However, research by Chen et al.~\cite{chen2023chatgpt} highlights that the direct use of LLMs, such as ChatGPT-4, for detecting smart contract vulnerabilities has produced very unsatisfactory results, a domain where traditional program analysis techniques excel ~\cite{chen2021defectchecker}. The strength of traditional program analysis techniques lies in their ability to enhance the comprehension of complex code structures through abstract data (e.g., control flow graph~\cite{allen1970control} and data flow graph~\cite{davis1982data}). This capability is often beyond the reach of LLM. In addition, LLM is susceptible to issues such as hallucinations and randomness~\cite{yao2023llm}, which can lead to decreased accuracy in vulnerability detection.

To confirm if smart contract developers use vulnerable code from \textit{SO} code snippets, we ran an online survey and received 74 valid responses. Our survey results show that 88.4\% of smart contract practitioners rely on Q\&A websites such as \textit{SO} to solve problems encountered during the development process. However, less than 20\% of these practitioners then conduct thorough security checks on the code they reuse from these \textit{SO} posts. This suggests that forum-sourced smart contract code snippets indeed present high potential security risks. According to the survey feedback, ``lack of support for direct code analysis on \textit{SO}'' is the main reason why developers do not apply existing tools to security analysis of code snippets, which highlights the importance of tools like \textsc{SOChecker}.


To address this major real-world issue of incomplete smart contract code snippet security analysis, we introduce \textsc{SOChecker}, a tool that combines the code completion capabilities of LLMs with traditional program analysis methods. 
\textsc{SOChecker} is the \textit{first} tool specifically designed to analyze fragmented smart contract code on Q\&A websites such as \textit{Stack Overflow} and able to detect nine common smart contract vulnerabilities listed in DASP10~\cite{dasp2018}, e.g., \textit{Reentrancy}, \textit{Access Control}, etc.~\cite{chen2021defectchecker}. 
\textsc{SOChecker} comprises two key components: a \textit{Code Completer} and a \textit{Vulnerability Detector}. For the former we employ the \textit{Llama2} model~\cite{touvron2023llama}, fine-tuned with a dataset of the top 1,000 smart contracts with the highest transaction volume from the Ethereum mainnet, to enhance code completion capabilities. While a LLM can successfully complete the semantics of the program, the code it produces may occasionally exhibit syntax issues, such as incompatible \textit{Solidity} versions or missing structural symbols. Hence, we developed scripts for automated version matching and code structure completion to address these issues.
After completing SO code snippets into compilable contracts, we use a conventional \textit{Vulnerability Detector} approach for security analysis. Specifically, we first construct a Control Flow Graph (CFG)~\cite{allen1970control} of the contract. Considering that we only aim to detect vulnerabilities in the original code snippets and code added by the LLM may introduce new vulnerabilities, we implement a program pruning strategy to remove paths generated by LLM code from the CFG. \textsc{SOChecker} is able to detect all nine vulnerabilities categorized by DASP10~\cite{dasp2018}, a widely recognized smart contract vulnerability list. 

\begin{table*}[t]
    \centering
        \caption{Top 9 smart contract vulnerabilities in DASP10 and their corresponding descriptions. }
        \resizebox{0.9\linewidth}{!}{
            \begin{tabular}{l|l}
                \hline
                \textbf{Vulnerability} & \textbf{Description} \\
                \hline
Reentrancy (RE)& Reentrant function calls make a contract to behave in an unexpected way \\
Access Control (AC)& Failure to use function modifiers or use of tx.origin \\
Arithmetic Issues (AI)& Integer over/underflows \\
Unchecked Return Values (URV) & call(), callcode(), delegatecall() or send() fails and it is not checked \\
Denial of Service (DoS) & The contract is overwhelmed with time-consuming computations \\
Bad Randomness (BR) & Malicious miner biases the outcome \\
Front Running (FR) & Two dependent transactions that invoke the same contract are included in one block \\
Time Manipulation (TM) & The timestamp of the block is manipulated by the miner \\
Short Address Attack (SAA) & EVM itself accepts incorrectly padded arguments \\
\hline
        \end{tabular}
        }
        \label{tab:types_of_vulnerabilities}
\end{table*}

To assess the efficacy of \textsc{SOChecker}, we curated a dataset of 897 Solidity code snippets from smart contract-related \textit{SO} posts. In the code completion stage, our fine-tuned model successfully completed 75.5\% of code snippets, outperforming the \textit{Llama2} base model, as well as widely used \textit{GPT-3.5-turbo}, and \textit{GPT-4} LLMs. In the vulnerability detection stage, our experimental results show that \textsc{SOChecker} achieved an F1 score of 83.4\%, greatly outperforming 10 state-of-the-art smart contract vulnerability detection tools.
 When applying \textsc{SOChecker} to analyze vulnerabilities in \textit{SO} code snippets directly,  it achieves an F1 score of 68.2\%, while the scores for \textit{GPT-3.5} and \textit{GPT-4} are only 20.9\% and 33.2\%, respectively.

In this research, we make the following key contributions:
\begin{itemize}[leftmargin=10pt]

\item We conducted a survey to collect the perspectives of smart contract practitioners on the usage of \textit{SO} code, demonstrating the high potential security risks associated with smart contracty code snippet usage from Q\&A websites.

\item We introduced \textsc{SOChecker}, the \textit{first} tool that combines code completion capabilities of LLMs with the program analysis methods to analyze smart contract code snippets found on Q\&A websites like \textit{Stack Overflow}. \textsc{SOChecker} is able to detect nine common smart contract vulnerabilities.

\item We curated a high-quality dataset consisting of 897 Solidity code snippets from smart contract-related posts on \textit{SO} posts. We used this dataset to evaluate \textsc{SOChecker}. The results indicate that the effectiveness of \textsc{SOChecker} is as high as 68.2\%, surpassing GPTs and other traditional vulnerability detection tools.

\item To promote further research in related fields, we make available our dataset, experimental results, and source code of \textsc{SOChecker} at \url{https://github.com/BugmakerCC/SOChecker}~\cite{SOChecker}. 
\end{itemize}


\section{Motivation and Background}
\label{sec:background}

\subsection{Motivating Example}
\label{subsec:motivating}
\begin{figure}[htbp]
\setlength{\abovecaptionskip}{0cm}
\setlength{\belowcaptionskip}{-0.3cm}
    \centering
    
    \includegraphics[width=3.3in]{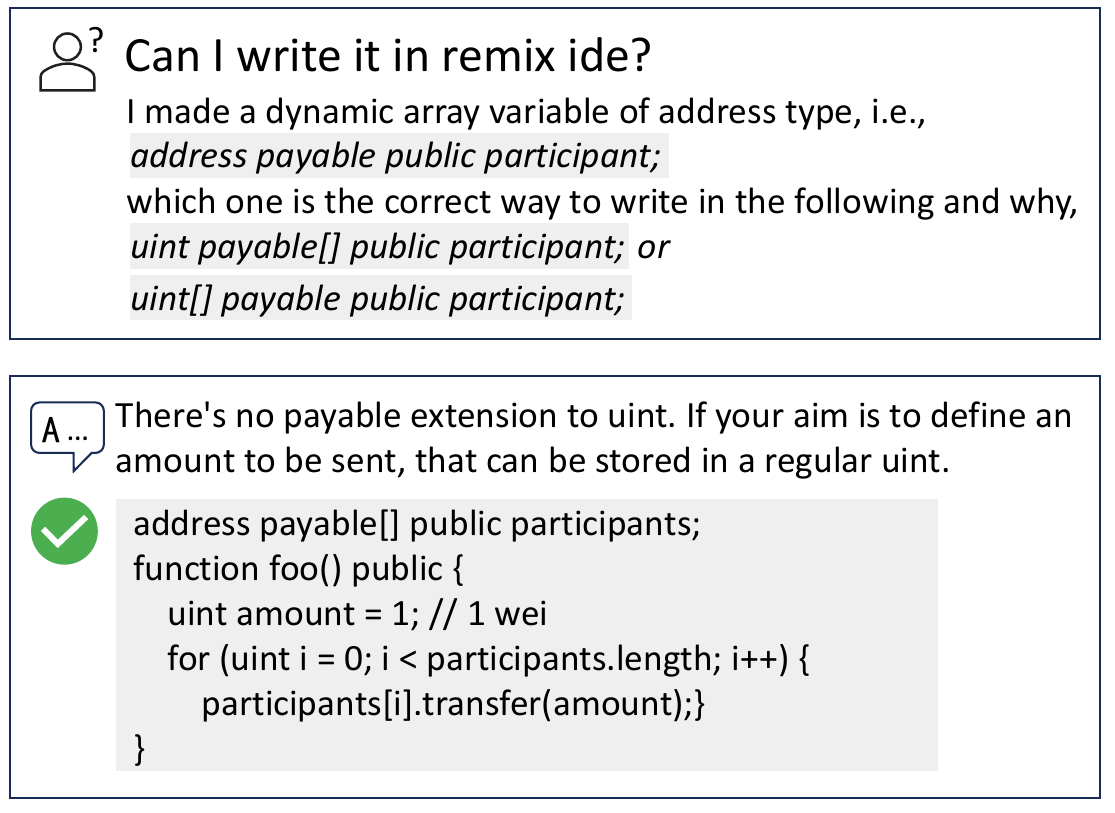}
    \caption{A post on \textit{Stack Overflow} related to \textit{Solidity}.}
    \label{fig:me}
    
\end{figure}

In Figure~\ref{fig:me}, we show a \textit{SO} post~\footnote{https://stackoverflow.com/questions/72171101/can-i-write-it-in-remix-ide} as an example, where
the questioner posted a question about a Solidity programming issue. In addition to answering the question, the respondent also provided a code snippet for reference. Although this code may serve the immediate needs of the inquirer -- evidenced by the acceptance of the answer -- it conceals a \textit{Denial of Service}~\cite{DoS} vulnerability hidden in it. This vulnerability could be exploited by a malicious attacker to disrupt the normal execution of this function, preventing legitimate participants from receiving payment. Even worse, this vulnerable code may also be reused by other developers reading this SO post, if they encounter similar issues.  If several people up-rate the post, it may become a popular solution despite the vulnerability. 


Detecting vulnerabilities in this code snippet is not easy. Firstly, like most such code snippers, the code in the post is fragmented and uncompilable, which cannot be straightforwardly analyzed using traditional contract vulnerability detection tools. 
Unlike conventional program analysis tools, LLMs can directly detect vulnerabilities in code snippets without requiring compilable code. However, Chen et al.~\cite{chen2023chatgpt} have shown that LLMs face challenges in directly detecting vulnerabilities in contract code, frequently resulting in false positives. Therefore, a more accurate and precise method is needed to analyze the security of smart contract code snippets.

\subsection{DASP10 Smart Contract Vulnerabilities}
\label{subsec:dasp10}

The DASP10~\cite{dasp2018} is a list of common smart contract vulnerabilities and is frequently referenced in academic research~\cite{chen2023chatgpt, FerreiraEtAl2020ASE}. Smartbugs~\cite{DurieuxEtAl2020ICSE}, a framework built upon the DASP10 vulnerability classification, integrates multiple vulnerability detection tools and has been employed in numerous subsequent studies~\cite{diAngeloEtAl2023EMSE,chen2023chatgpt}. A summary of each vulnerability listed in DASP10 is shown in Table~\ref{tab:types_of_vulnerabilities}. Only the top nine vulnerabilities are discussed, since the 10th vulnerability of DASP10 is ``Unknown Unknowns'', representing all undiscovered vulnerabilities. While some research~\cite{xblock} has outlined increasingly complex types of smart contract vulnerability, our study focuses primarily on code snippets sourced from \textit{Stack Overflow (SO)}. These snippets are typically brief, straightforward in logic, and generally free of intricate vulnerabilities. Therefore, we adopt DASP10 as the detection standard.

\subsection{Large Language Models}
A Large Language Model (LLM) is a machine learning model acquired through extensive training on a substantial corpus of text data. This allows it to comprehend and generate natural language and other textual formats proficiently~\cite{wei2022emergent,kasneci2023chatgpt}. Throughout their training process, most LLMs acquire extensive code knowledge from code-based training data. This results in high proficiency in code-related tasks, e.g., code generation, comprehension, and summarization\cite{hou2024large}. Research indicates a growing inclination among programmers to use LLMs to generate code to aid in work\cite{vaithilingam2022expectation}. The code completion ability of LLMs has also garnered recognition~\cite{eghbali2024hallucinator}. Furthermore, relevant studies have developed LLM-based code translation tools that have shown strong performance~\cite{pan2023stelocoder}.

\textit{GPT-4}~\cite{gpt4} is an LLM developed by \textit{OpenAI} specifically for natural language processing and text generation. As the latest iteration of the \textit{GPT} series, it builds on and refines the technological advances of its predecessor. In particular, surpassing \textit{GPT-3}, \textit{GPT-4} showcases substantial improvements in terms of model size, training data size, and overall performance. This model follows a closed-source approach with a commercial licensing model, which requires users to pay a fee for access rights~\cite{apiprice}. Despite the associated cost, \textit{GPT-4} demonstrates exceptional proficiency in a multitude of tasks, making it a widely acclaimed and popular LLM.
Recently, \textit{Meta} released their latest open-source large language model, \textit{Llama 2}~\cite{touvron2023llama}. Its pre-trained model is trained on 2 trillion tokens and its fine-tuning model has been trained on more than 1 million human annotations~\cite{llama2}. \textit{Llama 2} outperforms other open-source language models on many external benchmarks, including reasoning, coding, proficiency, and knowledge tests~\cite{llama2}. Moreover, \textit{Llama 2} is available for free research and commercial use~\cite{llama2}, so we can build datasets specific to a task and fine-tune it based on the model, making the fine-tuned model more capable of handling the task.

\subsection{Pre-training, Fine-tuning and Inference}
LLMs typically undergo pre-training and fine-tuning during their training process. In the pre-training phase, the model is exposed to extensive text data to acquire linguistic knowledge that includes grammar, context, and semantics~\cite{tirumala2023d4}. Following completion of pre-training, the model often undergoes fine-tuning to tailor its capabilities to specific tasks, e.g., code generation and summarization~\cite{sun2019fine}. The fine-tuning stage generally employs supervised learning~\cite{cunningham2008supervised}, utilizing labeled data for additional training to adjust model parameters and align with the requirements of the targeted tasks. 

Inference refers to the process in which a trained model generates output based on input data~\cite{abdelouahab2018accelerating}. This typically occurs when the model has completed training and is prepared to process real-world data. This step is crucial to applying the model to practical problems and tasks. LLMs suffer from instability (i.e., the responses generated each time are different) and hallucinations (i.e., the responses generated contain things that have not appeared in context) during inference due to sensitivity to adversarial samples\cite{wang2019assessing}, overfitting~\cite{hawkins2004problem}, and complex context.

\section{Real-world \textit{Stack Overflow} Smart Contract Code Snippet Usage}
\subsection{Motivation}
While smart contract codes on \textit{Stack Overflow (SO)} may contain vulnerabilities, it remains unclear whether these codes are actually utilized by developers in practice. We conducted an online survey specifically targeting real-world smart contract developers to try and determine how often this actually happens. Our survey was designed to gather insights on smart contract developer perspectives and usage of code snippets from \textit{SO}. This includes their criteria for using code snippets, approaches to evaluate and modify code quality and security before integrating them into their projects. Based on the information, we can assess the risks of using such community-contributed code.

\subsection{Survey Design}
We followed Kitchenham and Pfleeger’s instructions~\cite{kitchenham2008personal} for personal opinion surveys and designed an anonymous survey to increase response rates~\cite{tyagi1989effects}. Our survey was made available in both English and Chinese, since English is the most widely used language and Chinese has the largest number of speakers worldwide. Bi-lingual co-authors carefully reviewed the two versions and guaranteed their consistency. For each question, we made it an optional question to prevent practitioners from not understanding it or being unwilling to answer it. The following provides a brief introduction to our survey questions. For the complete questionnaire, please refer to our online supplementary material~\cite{SOChecker}.

\noindent \textbf{Demographics ($Q_{1-4}$).} We collected the following demographic information to understand the background of respondents, and filter those who might not fully understand our survey.
\begin{itemize}
    \item Smart contract practitioner? \textit{Yes / No.} ($Q_{1}$)
    \item Main role as a smart contract practitioner. \textit{Development / Testing / Project Management / Research / Other.} ($Q_{2}$)
    \item Experience in years. \textit{Free-text.} ($Q_{3}$)
    \item Current country of residence. \textit{Free-text.} ($Q_{4}$)
\end{itemize}

\noindent \textbf{Access Frequency of \textit{Stack Overflow}} ($Q_{5-6}$). We wanted to ask practitioners to self-assess their familiarity with \textit{SO} ($Q_{5}$), and how often they access the Q\&A platform ($Q_{6}$). 

\noindent \textbf{Questioners' Perspective ($Q_{7-12}$).}  We wanted to examine how practitioners, acting as questioners, perceive and engage with smart contract codes on \textit{SO}. We first assessed how frequently these practitioners ask questions on \textit{SO} ($Q_{7}$), and the types of questions related to smart contracts (such as grammar, security issues, API uses, etc) that most concern them ($Q_{8}$). Additionally, we asked whether practitioners used code directly from \textit{SO} ($Q_{9}$). For those who have used SO code snippets, we asked about the security analysis they performed prior to code incorporation into their own smart contract programs ($Q_{10}$). Conversely, for practitioners who have never used \textit{SO} code snippets, we asked for their reasons ($Q_{11}$). 
Finally, we askded about the security assessment measures, e.g., code reviews, that practitioners adopt when evaluating code from \textit{SO} ($Q_{12}$).

\noindent \textbf{Respondents' Behaviour ($Q_{13-15}$).} 
We asked about how often practitioners respond to others' questions on \textit{SO} ($Q_{13}$). Subsequently, we asked respondents if they verify the security of any code snippets they add to their answers ($Q_{14}$), and if so,  the methods they use to ensure code security before sharing it ($Q_{15}$). 

\noindent \textbf{Understanding of Smart Contract Vulnerabilities ($Q_{16-18}$).}  
We showed a set of example smart contract code vulnerabilities as outlined in DASP10~\cite{dasp2018}, inviting practitioners to explain both their understanding ($Q_{16}$) and their perception on the importance of identifying these vulnerabilities in \textit{SO} code ($Q_{17}$). Additionally, we asked practitioners to suggest additional contract vulnerabilities that they consider necessary to detect ($Q_{18}$).

\noindent \textbf{Usage of tools  ($Q_{19-21}$).} 
We asked practitioners about the use of existing smart contract vulnerability detection tools. First, we investigated how often practitioners utilize these tools in their development process ($Q_{19}$), and whether they apply these tools for \textit{SO} code ($Q_{20}$). Then, we asked practitioners' perspectives on the main limitations associated with employing these tools for security analysis of \textit{SO} code ($Q_{21}$).

\noindent \textbf{Suggestions for Improvement ($Q_{22}$).}  Finally, we asked what aspects do practitioners think need improvement when reviewing or using community-shared smart contract code snippets ($Q_{22}$).

\subsection{Survey Validation} We conducted a pilot survey with a small number of practitioners to obtain feedback on whether the questions are clear and easy to understand. The participants included our academic collaborators and partners working in well-known blockchain companies. Based on the feedback, we refined some questions for enhanced clarity without adding or removing any questions. We also polished our translation to further reduce ambiguity between the two language versions of the survey.
\subsection{Participant Recruitment} We adopted a non-probabilistic~\cite{gabor2007types} strategy for participant recruitment. Specifically, we conducted a keyword-based search for smart contract repositories on Github, extracted their contributors’ emails via the Github REST API~\cite{github2023restapi}, and sent the survey to them. Our selection of keywords encompasses a broad spectrum of topics within smart contract technology, e.g., ``smart contract'', ``solidity'', and ``erc-20''. For the complete keyword list, please refer to our online repository~\cite{SOChecker}.
We then sent our survey to a total of 1,416 smart contract practitioners and received 74 valid responses from 19 countries, a reasonable number compared to previous smart contract related survey~\cite{zhang2023contracts, wan2021smart, bosu2019understanding}. We  excluded five responses, as they claim to have no development experience in smart contracts. The roles played by respondents in the field of smart contracts are mainly distributed in research (31, 44.9\%), development (52, 75.4\%), testing (33, 47.8\%), project management (12, 17.4\%), security audit (37, 53.6\%), compliance check (3, 4.3\%), training and education (9, 13.0\%) and market analysis (2, 2.9\%). Their average years of experience are 2.80 (min: 0.2, max: 7.0, median: 2.0, sd: 2.0).

\subsection{Results}

\noindent {\bf{Access Frequency of \textit{Stack Overflow}.}} Only 8.1\% participants identified themselves as either ``Ignorant'' or ``Not very familiar'' with \textit{SO}, suggesting that a majority of 91.9\% possess some degree of familiarity with the platform. Notably, the vast majority (94.6\%) of practitioners have accessed \textit{SO}, and 77.0\% of the participants reported engaging with \textit{SO} at least once a week. These findings highlight the important role of \textit{SO} in the smart contract ecosystem.

\begin{figure}[htbp]
\setlength{\abovecaptionskip}{0.5cm}
\setlength{\belowcaptionskip}{0cm}
    \centering
    \includegraphics[width=\columnwidth]{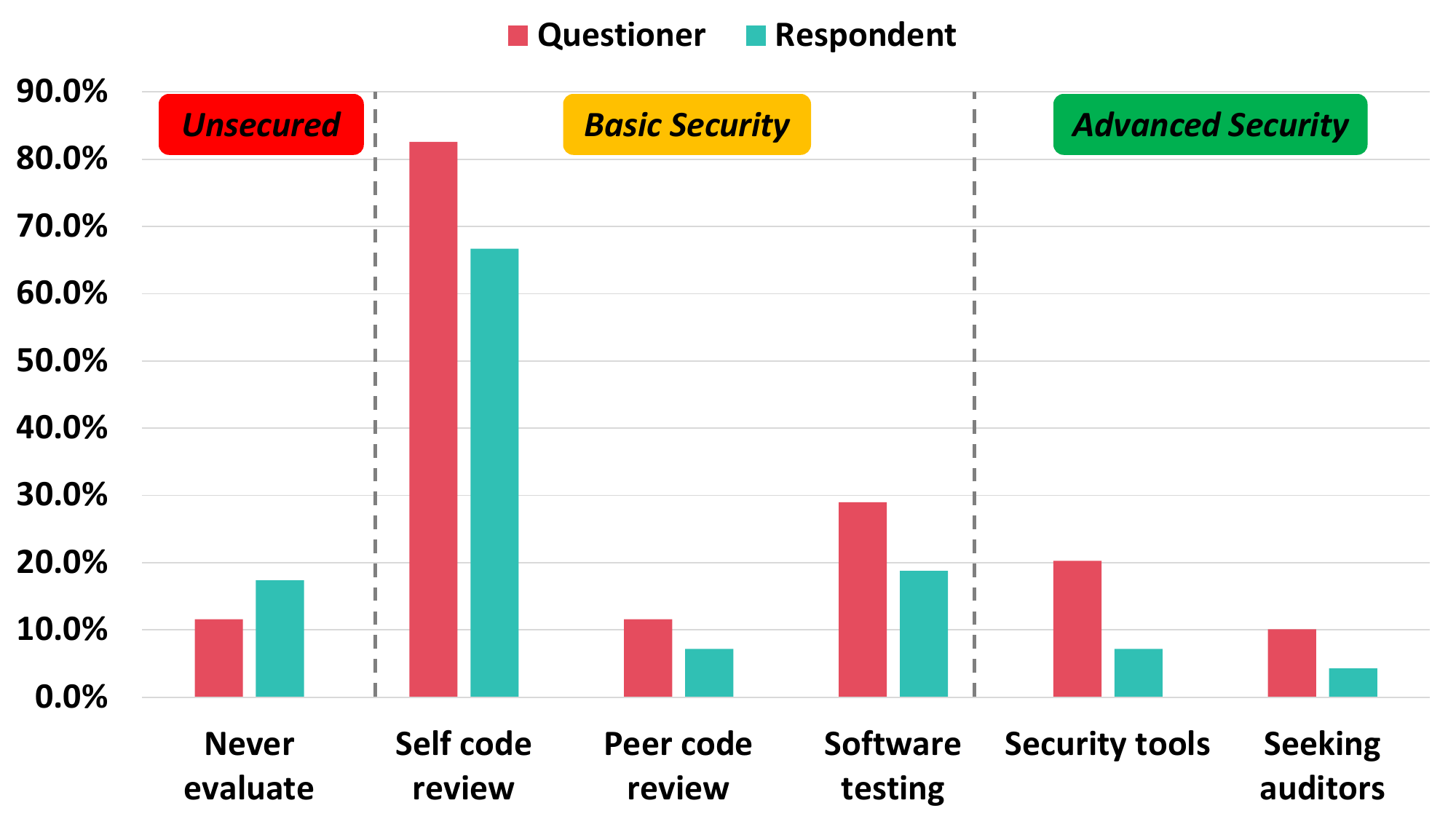}
    \caption{The way for participants to conduct security analysis on the code on \textit{Stack Overflow}.}
    \label{fig:sec_ana}
\end{figure}

\noindent {\bf{Questioners’ Perspective.}} Although 59.4\% of our participants infrequently post questions on \textit{SO}, a majority have used code from the site (88.4\%). Before using code from \textit{SO}, only 16.4\% performed comprehensive security audits using various methods. Over a third perform basic code reviews without employing additional tools or methods for security audits (36.1\%). Many said they understand the code logic but do not specifically assess code security (31.1\%). Few (11.6\%) of the participants indicated that they do not refer to the code from \textit{SO}, the predominant reason being that the code does not align with the unique requirements of their projects (60.0\%). Figure ~\ref{fig:sec_ana} shows the proportion of participants conducting security analysis on code from \textit{SO} in various ways. We categorize the security status of the code into three distinct levels, determined by the rigor of the auditing methods employed. These levels are defined as ``unsecured'', ``basic security'' and ``advanced security''. Most participants favor self-review to identify obvious errors and vulnerabilities (82.6\%). In contrast, the use of more professional audit methods, such as specialized security audit tools (20.3\%) or consulting professional auditors (10.1\%), is significantly less common.


\begin{center}
    \begin{myboxc} \textbf{Observation 1:} The vast majority (94.6\%) of practitioners have accessed \textit{SO}. Although frequent questioning by practitioners on the platform is rare, many (88.4\%) seek solutions by browsing through posts made by others.
    \end{myboxc}
\end{center}

\noindent {\bf{Respondents’ Behaviour.}} When responding to queries on \textit{SO}, only a very small proportion of respondents (7.1\%) consistently verify the security of the code before providing any code-related responses, regardless of the context. A larger group conduct security checks only for complex codes or those involving sensitive functions such as transfers (28.6\%). Approximately a quarter (25.7\%) check the security of the code in most instances. Almost a quarter (24.3\%) of our respondents never evaluate security of their example code before responding to others' questions with it. The most prevalent method employed by respondents for checking code security is through self-review (65.7\%). Only a very small fraction utilize professional tools (7.1\%) or seeks professional assistance (4.3\%) for security analysis. We determined the percentage of participants capable of ensuring advanced security of shared code snippets, and discovered that this group represents merely 13.6\% of the total.


\begin{center}
    \begin{myboxc} \textbf{Observation 2:} Both as questioners and respondents, the majority of participants are only able to ensure basic even lower security for code (86.4\%), with a very limited number (13.6\%) capable of guaranteeing advanced security.
    \end{myboxc}
\end{center}

\begin{figure}[htbp]
\setlength{\abovecaptionskip}{0cm}
\setlength{\belowcaptionskip}{0cm}
    \centering
    \includegraphics[width=\columnwidth]{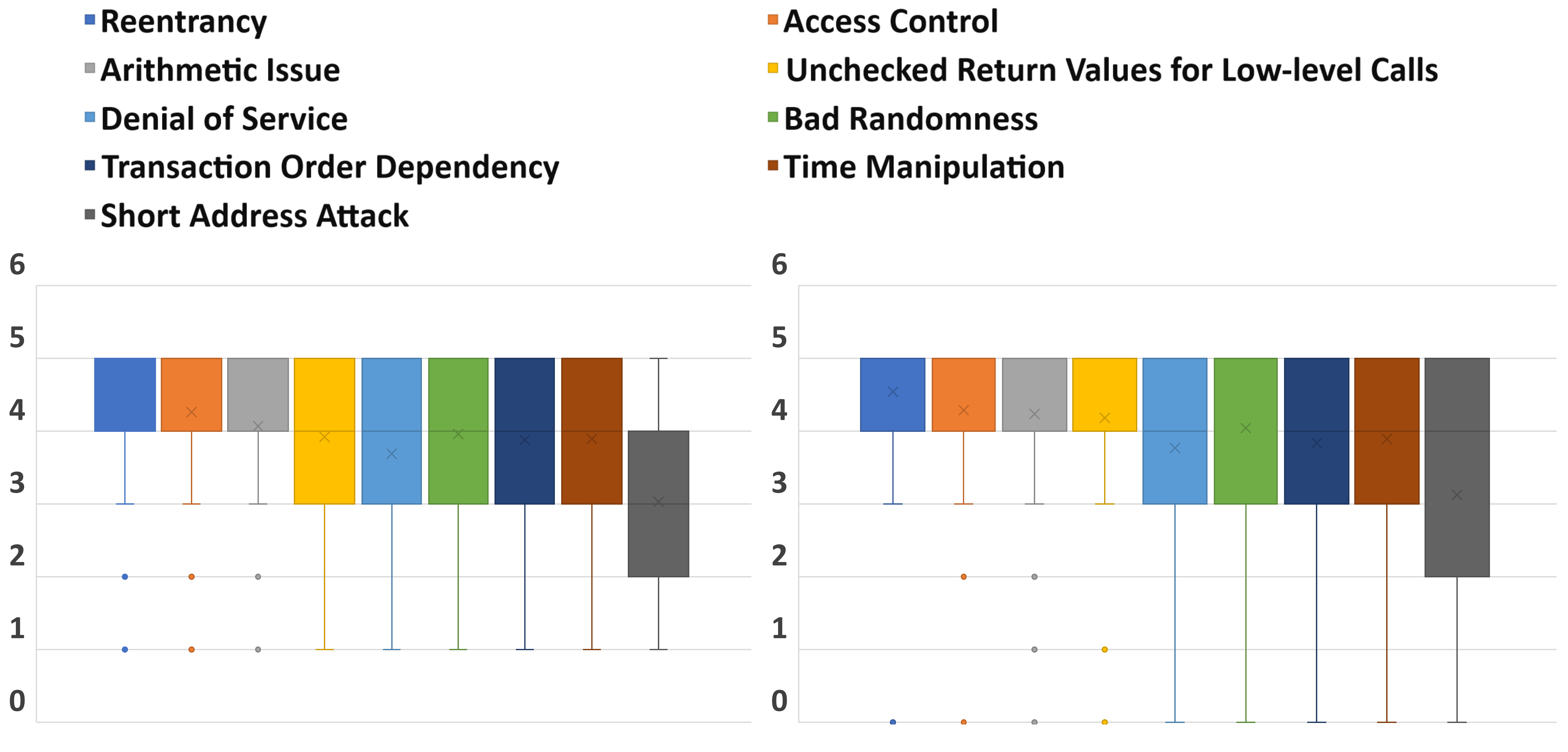}
    \caption{The level of understanding of smart contract vulnerabilities among participants (left) and their belief in the necessity of detecting these vulnerabilities on \textit{SO} (right).}
    \label{fig:boxplot}
\end{figure}

\noindent {\bf{Familarity with Smart Contract Vulnerabilities.}} Figure~\ref{fig:boxplot} illustrates participants' comprehension of smart contract vulnerabilities and their perceived importance of detecting these vulnerabilities on \textit{SO}. Participants assigned scores ranging from 1 to 5 for each vulnerability, with 1 being the lowest and 5 the highest familarity. Among the vulnerabilities listed in DASP10~\cite{dasp2018}, participants demonstrated the highest level of understanding regarding \textit{Reentrancy}, with an average score of 4.30. On the contrary, their understanding of \textit{Short Address Attack} was the lowest, averaging at 3.03. Concurrently, \textit{Reentrancy} is also perceived by participants as the most critical vulnerability to detect, receiving an average importance score of 4.54. Beyond the vulnerabilities outlined in DASP10~\cite{dasp2018}, participants also identified additional concerns, including price manipulation and accuracy issues.

\noindent {\bf{Usage of tools.}} A significant majority of participants (81.4\%) have utilized tools for detecting vulnerabilities in smart contracts. Yet, only a very small 20.0\% have applied these tools for the security analysis of code on \textit{SO}. When questioned about their reluctance to use these tools on \textit{SO} code, 44.9\% cited ``lack of support for direct code analysis on \textit{SO}'' as a key factor, while 60.7\% pointed to ``poor usability of the tools, their complexity, and the high time cost involved'' as their primary concerns.


\noindent {\bf{Expectations for \textit{Stack Overflow}.}} For smart contract codes shared within the community, a significant proportion of our survey participants (72.9\%) view security and vulnerability detection as the primary areas in need of improvement or support. Additionally, there is a notable demand for improvements in code quality and clarity, as indicated by 64.3\% of the participants.

\begin{center}
    \begin{myboxc} \textbf{Observation 3:} Despite the availability of numerous tools for smart contract security, only 20.0\% of participants reported using these tools on code from \textit{SO}. The predominant reason cited for not using these tools, mentioned by 44.9\% of respondents, was the ``lack of support for direct code analysis on \textit{SO}''.
    \end{myboxc}
\end{center}


\section{Our \textsc{SOChecker} Approach}
\label{sec:method}

\subsection{Overview}
\begin{figure}[htbp]
\setlength{\abovecaptionskip}{0.1cm}
\setlength{\belowcaptionskip}{-0.5cm}
    \centering
    \includegraphics[width=\columnwidth]{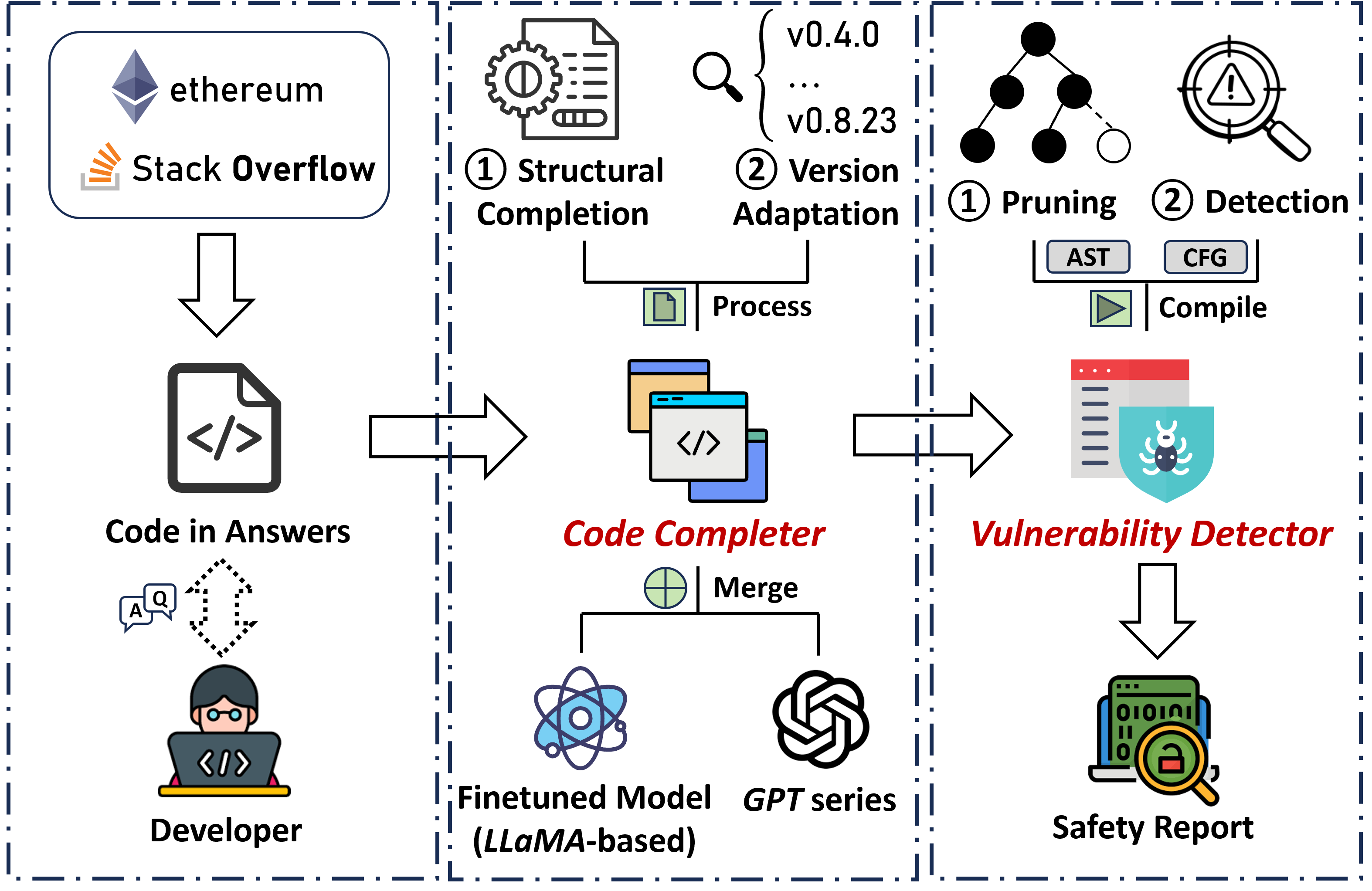}
    \caption{The overall workflow of \textsc{SOChecker}.}
    \label{fig:overview}
\end{figure}

Figure~\ref{fig:overview} provides an overview of \textsc{SOChecker} approach. \textsc{SOChecker} comprises two main components: a \textit{Code Completer} and a \textit{Vulnerability Detector}. For the \textit{Code Completer}, we first collect the top 1,000 smart contracts with high transaction volumes from the Ethereum mainnet. Then, we use them to fine-tune the open-source LLM \textit{llama2-chat-13b}~\cite{llama2}, aiming to enhance the model's performance in smart contract code completions.
For code snippets that our model cannot complete, we further utilize \textit{GPTs} to generate completions again and then merge the results. Finally, we performed two steps (i.e., structural completion, version adaptation) on the LLM-completed code to increase the number of compilable smart contracts.
Our \textit{Vulnerability Detector} utilises two primary steps, code preprocessing and vulnerability detection.
During the code preprocessing stage, we compile the completed smart contract code, extract their Abstract Syntax Trees (ASTs), and construct Control Flow Graphs (CFGs). Subsequently, we prune the CFGs based on the original code snippets and ASTs. In the vulnerability detection stage, we developed  patterns for nine types of DASP10 vulnerabilities and conduct pattern matching on the pruned CFG to identify potental vulnerabilities. Based on the results of vulnerability detection, we generate a safety report for developers to consult.

\subsection{Code Completer}
\subsubsection{Data Collection}
\label{sec:data_collection}
\label{data_collection} To construct a smart contract dataset for fine-tuning, we first gathered 345,058 open-source smart contracts from the Ethereum mainnet through the GitHub repository \textit{smart-contract-sanctuary}~\cite{sanctuary} up to March 2023. 

To evaluate the code completion performance of the all the candidate models (i.e., \textit{GPTs}, \textit{llama2}, \textit{codellama}), we obtained 4,952 posts related to smart contracts from \textit{Stack Overflow (SO)} after October 2021~\footnote{The training data for the \textit{GPTs} was up to September 2021~\cite{gpt4}. Selecting posts after this time point helps mitigate the impact of data leakage.}. These posts were selected based on the following criteria: 1) The post should have at least one response. 2) The post should include at least one tag of \textit{Solidity}, \textit{Ethereum}, \textit{ERC20}, \textit{ERC721}, or \textit{Contract}. From these 4,952 posts, we refined our dataset by excluding answers for several reasons: 1) Non-code answers were removed, as our analysis focuses find security issues on code snippets. 2) Code  was not written in the \textit{Solidity} language was omitted. 
3) Brief one-liner code snippets were omitted, as they typically provide limited information and are unlikely to pose significant security risks. Finally, we had a curated dataset comprising 897 code snippets for our subsequent code completion and analysis steps.

\subsubsection{Data Preprocessing \& Fine-Tuning}
\label{sec:finetune}
\begin{figure}[htbp]
\setlength{\abovecaptionskip}{0cm}
\setlength{\belowcaptionskip}{-0.5cm}
    \centering
    \includegraphics[width=\columnwidth]{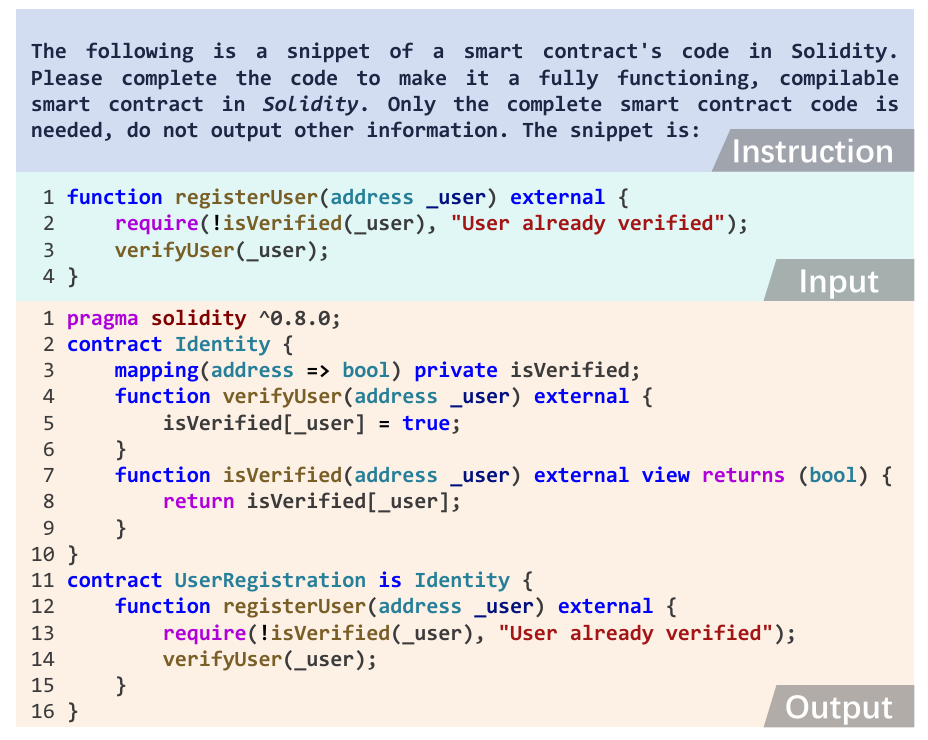}
    \caption{An example of constructing fine-tuning data.}
    \label{fig:finetune}
\end{figure}

We adopt a fine-tuning process to enhance the model's ability to accurately understand and generate smart contract code. We strategically chose to focus on a subset of the top 1,000 smart contracts with the highest transaction volumes from the initial pool of 345,058 smart contracts for fine-tuning. The average length of these 1,000 smart contracts is 711.03 lines, including comments. Choosing these smart contracts for fine-tuning is to balance the breadth of smart contract applications (e.g., NFTs, DeFi) with the practicality of computational efficiency during fine-tuning. Meanwhile, the selection with the highest transaction volume is based on the assumption that these contracts are more likely to represent real-world scenarios with significant usage and functionalities. Additionally, these smart contracts typically present lower vulnerability risks. 
From collected data, we found that the majority of code snippets on \textit{SO} are function-level segments. Therefore, function-level code was used as input for LLM fine-tuning, with complete codes serving as the targeted outputs. We construct fine-tuning data by extracting functions from complete contracts.
 However, incorporating lengthy smart contracts posed a challenge; their complexity and detailed nature could potentially lead the model to be distracted by the intricacies of the logic itself, thus affecting the effectiveness of the model. 
To mitigate this issue, we used a method to segment lengthy contracts into shorter parts. 
Our segmentation process starts with the compilation of each smart contract to obtain its abstract syntax tree (cf. Section~\ref{subsubsec:codeprepro}), providing insights into the dependency relationships among various subcontracts. For each subcontract, we use its function snippets as the input of a fine-tuning data and include the subcontracts upon which it depends, along with itself, as the output of a fine-tuning data. 
Figure~\ref{fig:finetune} illustrates a simple example of constructing fine-tuning data. The function \textit{registerUser} on line 1 of ``Input'' serves as the target objective that requires completion. Given that this function is dependent on other contracts, we include the subcontract \textit{UserRegistration} on line 11 of ``Output'' containing this function and its dependent contract \textit{Identity} on line 2 of ``Output'' as the completed code. The task instructions are further detailed in Figure~\ref{fig:finetune}. This procedure allows us to construct a dataset of segmented smart contracts for fine-tuning effectively.

\subsubsection{Model Selection}
In our study, we chose \textit{llama2-chat-13b}~\cite{llama2} and \textit{Codellama-instruct-13b}~\cite{roziere2023code} for fine-tuning with the following reasons: 1) Both of them are open-source, ensuring transparency and accessibility. 2) These LLMs possess a moderate number of parameters~\cite{li2024eagle}, resulting in an acceptable computational cost. 3) Both can be fine-tuned to better understand task requirements. 4) Previous research and relevant work~\cite{roumeliotis2023llama,cifarelli2023safurai,alrashedy2023language} have demonstrated their effectiveness in various contexts. After fine-tuning, we compared their performance in the code completion task and then selected the model with better performance for further analysis.

We used the dataset constructed in Section~\ref{sec:data_collection} to fine-tune \textit{Llama2-chat-13b} and \textit{Codellama-instruct-13b}. We employed LoRA technology~\cite{hu2021lora} for model fine-tuning. The training parameters were configured with a rank of 8, alpha of 32, batch size of 256, 3 epochs, and a learning rate of 1e-4. All parameter settings remain default. To assess their performance, we randomly selected 324 smart contracts based on a confidence interval of 10 and a confidence level of 95\%~\cite{calculator} from the \textit{smartbugs-wild} dataset~\cite{DurieuxEtAl2020ICSE}, which comprises 47,398 smart contracts. Each smart contract's functions were extracted as code snippets for completion with the same method introduced in Section~\ref{sec:finetune}. In particular, these code snippets do not overlap with the fine-tuning dataset. The experimental results reveal that the fine-tuned \textit{llama2-chat-13b} provides 141 compiled smart contracts, whereas the fine-tuned \textit{Codellama-instruct-13b} only provides 48. This indicates that \textit{llama2-chat-13b} serves as a more suitable base model for fine-tuning in the task of smart contract code completion.

\subsubsection{Structural Completion}
Smart contract programming demands high syntactical precision, including the accurate placement of structural symbols, e.g., `)', `\}'. LLMs may struggle with the correct prediction of such symbols due to complex contextual relationships, often generating non-compilable code. 
To mitigate this issue, we implemented a preprocessing step for the LLM-generated code and developed a script to intelligently insert the missing structural symbols, thus ensuring the syntactical completeness and integrity of the code.

\subsubsection{Solidity Version Adaptation}
The version declaration of the smart contract is susceptible to errors. We observed instances where the code for certain contracts, despite being correct, failed compilation, with the issue solely attributed to the version declaration of \textit{Solidity}. This challenge arises from the numerous versions of \textit{Solidity} and the subtle differences between each version, making it challenging for LLMs to accurately discern the correct version from their extensive learned knowledge. To address this issue, we disregarded the version numbers generated by LLMs. Instead, we implemented scripts to systematically test and compile smart contracts across all \textit{Solidity} versions, thereby ensuring the compatibility and successful compilation.

\subsection{Vulnerability Detector}

\begin{figure}[htbp]
\setlength{\abovecaptionskip}{0cm}
\setlength{\belowcaptionskip}{-0.4cm}
    \centering
    \includegraphics[width=\columnwidth]{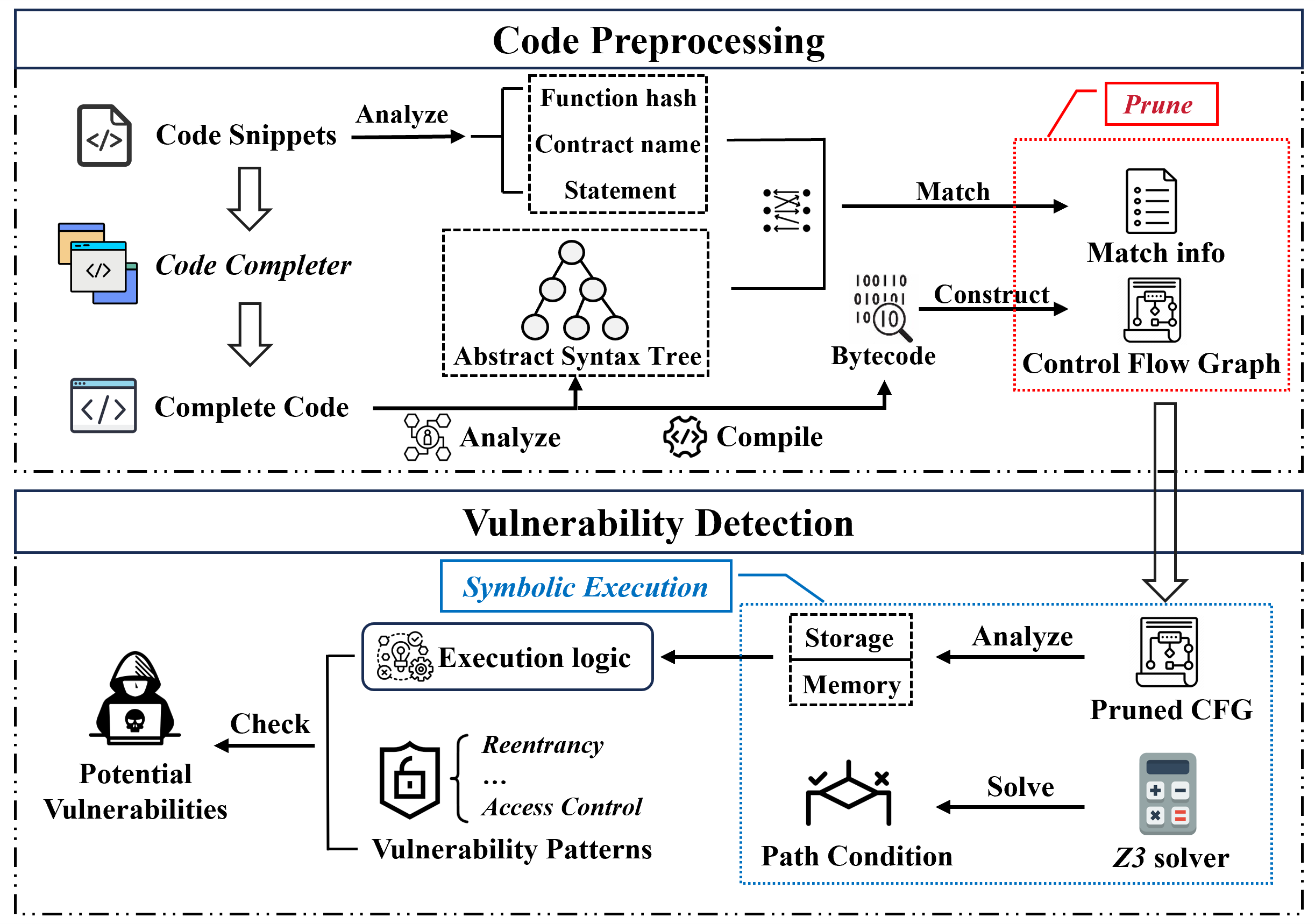}
    \caption{The overall workflow of \textit{Vulnerability Detector}.}
    \label{fig:vul_detector}
\end{figure}

\subsubsection{Code Preprocessing}
\label{subsubsec:codeprepro}

The complete smart contract code undergoes compilation by the \textit{Solidity} compiler, resulting in the generation of the corresponding Abstract Syntax Tree (AST)~\cite{neamtiu2005understanding} and bytecode. The source code information contained within the AST can be utilized for subsequent pruning steps. For bytecode, we utilize the API offered by Geth~\cite{geth} to disassemble it, enabling us to obtain the corresponding opcodes. Subsequently, we segment the opcodes into basic blocks and facilitate block-to-block jumps to finalize the construction of the Control Flow Graph (CFG)~\cite{allen1970control}.

Before performing program analysis, we prune the CFG of the smart contract to concentrate solely on original code fragments. While LLMs are capable of expanding code fragments into full smart contracts, the LLM-generated code might introduce bugs or vulnerabilities not in the original SO code snippet itself. Our pruning approach thus mitigates the influence of LLM-generated code and decreases the probability of false positives. The pruning algorithm is shown in Algorithm ~\ref{alg:pruning}. We first compile the smart contract and obtain its AST. Then, we extract some key information (e.g., function names and contract names) from the original code snippet through regular expression matching and initialize an empty list $subgraphs$ to store the subgraph related to the original code snippet in the AST. Next, we traverse each node in the AST to determine whether the information of that node matches the information in $info$. If so, we extract the subgraph where the node is located and add $subgraph$ to $subgraphs$. Finally, we merge and store all subgraphs in $prunedAST$, and extract the corresponding CFG based on the information of each node in $prunedAST$. This will result in a pruned CFG.

\begin{algorithm}
\caption{Pruning of smart contract snippet}\label{alg:pruning}
\begin{algorithmic}[1]
\Require \textbf{Contract} $ctr$, \textbf{Snippet} $snp$
\Ensure Pruned $CFG$
\State $ast \gets GetAST(ctr)$ 
\State $info \gets ExtractInfo(snp)$ 
\State $subGraphs \gets []$ 
\For{$node$ in $ast$}
    \If{$NodeInfo(node)$ in $info$} 
        \State $subGraph \gets ExtractSubGraph(node)$ 
        \State $subGraphs.append(subGraph)$ 
    \EndIf
\EndFor
\State $prunedAST \gets SubgraphMerging(subGraphs)$ 
\State $prunedCFG \gets ExtractCFG(prunedAST)$ 
\State \textbf{return} $prunedCFG$
\end{algorithmic}
\end{algorithm}


It should be noted that pruning may fail for the following two reasons: 1) The original code snippet lacks sufficient information to extract statement-level or higher details from the AST, leading to unsuccessful matching. 2) LLMs incorrectly completed the original code (e.g., changed function names,  deleted parts of the snippet source code), resulting in failure to retrieve any information about the original code snippet from the AST.

\subsubsection{Vulnerability Detection.}
After obtaining the pruned control flow graph, we carry out vulnerability detection on the original code snippet. Given that the code snippets found on \textit{SO} are typically simple, it is less likely we will encounter complex vulnerabilities, e.g., a price manipulation attack~\cite{kong2023defitainter}. Consequently, our analysis concentrates on vulnerabilities as categorized by the Decentralized Application Security Project (DASP) Top 10~\cite{dasp2018}. 


Numerous studies have defined patterns associated with these DASP10 vulnerabilities~\cite{oyente,smartcheck,slither,securify}. Building on their efforts, we have encapsulated these patterns within the program's CFG to facilitate vulnerability detection via symbolic execution. \textit{Z3} SMT solver~\cite{de2008z3} was used in \textit{Vulnerability Detector}. Taking \textit{Denial of Service (DoS)} as a case in point,  our initial step involves analyzing the CFG for loops, which suggests the presence of loops between program blocks. If no loop is present, exit; otherwise, proceed. We iterate through all instructions within blocks of the loop, identifying any resource-intensive operations such as `CALL' associated with functions like \textit{call} and \textit{transfer} in the code. Such functions typically involve substantial gas consumption. If these instructions are found, we conclude that a \textit{DoS} vulnerability has been detected. 
For details of each vulnerability pattern please consult our repository~\cite{SOChecker}.

\section{\textsc{SOChecker} Evaluation}
\label{sec:evaluation}
We conducted a detailed empirical evaluation of \textsc{SOChecker}, focusing on answering three key research questions: 
\begin{itemize}[leftmargin=9pt]
    \item RQ1: How effective is \textsc{SOChecker}'s \textit{Code Completer} in code completion of smart contract snippets?
    \item RQ2: How effective is \textsc{SOChecker}'s \textit{Vulnerability Detector} in vulnerability detection in completed, pruned smart contract code?
    \item RQ3: How does \textsc{SOChecker} perform when detecting vulnerabilities in real code snippets from \textit{Stack Overflow (SO)}?
\end{itemize}
For RQ1, we assessed the effectiveness of the \textit{Code Completer} on 897 code snippets collected from \textit{SO} (details see~\ref{sec:data_collection}). 
We evaluated both the compilability and correctness of all completed code, resulting in a LLM-completed smart contract dataset that was accurately completed by LLM. For RQ2, we evaluated the performance of our \textit{Vulnerability Detector} using the LLM-completed smart contract dataset, 
which is also compatible with traditional vulnerability detection tools due to the compilability of the code in the dataset. This approach enables a fair comparison with other traditional vulnerability detection tools. For RQ3, we conducted a comprehensive evaluation of \textsc{SOChecker}'s performance on the entire dataset and compared it with the performance of other LLMs (i.e., GPT-3.5-turbo and GPT-4). By addressing these three research questions, we aim to comprehensively evaluate \textsc{SOChecker}'s capabilities in 
detecting vulnerabilities in real code snippets from \textit{SO}.

\subsection{RQ1: Effectiveness of \textit{Code Completer}}
\label{completion}

We evaluated the effectiveness of \textit{Code Completer} by applying our fine-tuned model to the 897 real code snippets from \textit{SO}, as detailed in Section~\ref{data_collection}. When conducting the code completion process with our fine-tuned model, we observed a degree of uncertainty in the model's output, with some code snippets fail to be completed during the first iteration but successfully handled in the subsequent iterations. To fully leverage the model's capabilities, we employed multiple iterations of the code completion process. All the parameters of models (e.g., temperature, decoding strategy) remain default.

\noindent{\bf{Total Compilable Code.}} As shown in Figure~\ref{fig:model_comparison}, all the models have undergone 13 rounds of iteration, resulting in a continuous increase in the total number of successfully compiled smart contracts, reaching a plateau after 13 iterations. After 13 iterations, our fine-tuned model performs the best, indicating that our fine-tuning is effective for this task. Interestingly, the final performance of \textit{GPT-3.5-turbo} is slightly better than that of \textit{GPT-4}, which we believe is normal because although \textit{GPT-4} is a new version released after \textit{GPT-3.5-turbo}, it may also use updated training data and methods, which can lead to better performance on certain tasks while lowering it on others. In addition, the \textit{GPT} series models showed excellent performance at the beginning, and after the first round of code completion, they provide more than half of the compilable smart contracts. However, as the number of iterations continues to increase, the performance of the \textit{GPT} models decrease significantly. On the contrary, our model's performance has consistently improved, and after 13 iterations, it provides more compilable smart contracts than all other models.

\begin{table}[htbp]
\rowcolors{2}{gray!25}{white}
\setlength{\abovecaptionskip}{0cm}
\setlength{\belowcaptionskip}{-0.3cm}
    \begin{center}
        \caption{The code completion performance of models iterating on datasets.}
        \label{tab:iterating}
        \resizebox{\columnwidth}{!}{
        \begin{tabular}{c|c|c|c|c}
            \hline
            \rowcolor{gray!50} 
            \textbf{Models} & \textbf{\# Compilable} & \textbf{\# Correct} & \textbf{Time} & \textbf{Price} \\
            \hline
            \textit{Base} & 537 & 442 & \textbf{4.4h} & - \\
            \textit{GPT-3.5-turbo} & 774 & 669 & 7.8h & \$8.27 \\
            \textit{\textit{GPT-4}} & 736 & 649 & 11.2h & \$57.88 \\
            \textit{Ours} & 795 & 677 & 6.3h & - \\
            \textit{Ours+\textit{GPT}} & \textbf{889} & \textbf{846} & 7.1h & \$1.32 \\
            \hline
        \end{tabular}
        }
    \end{center}
\end{table}
\noindent{\bf{Quality of Compilable Code.}} 
We found that some models produced compilable code but deviated from expected behavior, attributable to design and training characteristics of LLMs. For example, LLMs may automatically repair vulnerabilities when completing the code or may change the logic of the original code. This behavior may stem from exposure to data and rules during LLM's training process, or may result from the model's biased interpretation of tasks or contextual understanding. Consequently, only assessing the volume of compilable code is insufficient -- the correctness of the model-completed code is equally important. We manually analyzed the code completed by various models. Table~\ref{tab:iterating} presents an evaluation of this metric. Our \textit{Code Completer} outperforms other LLMs in both the quantity and quality of compilable contracts.


\begin{figure}[htbp]
\setlength{\abovecaptionskip}{0cm}
\setlength{\belowcaptionskip}{-0.5cm}
    \centering
    \includegraphics[width=\columnwidth]{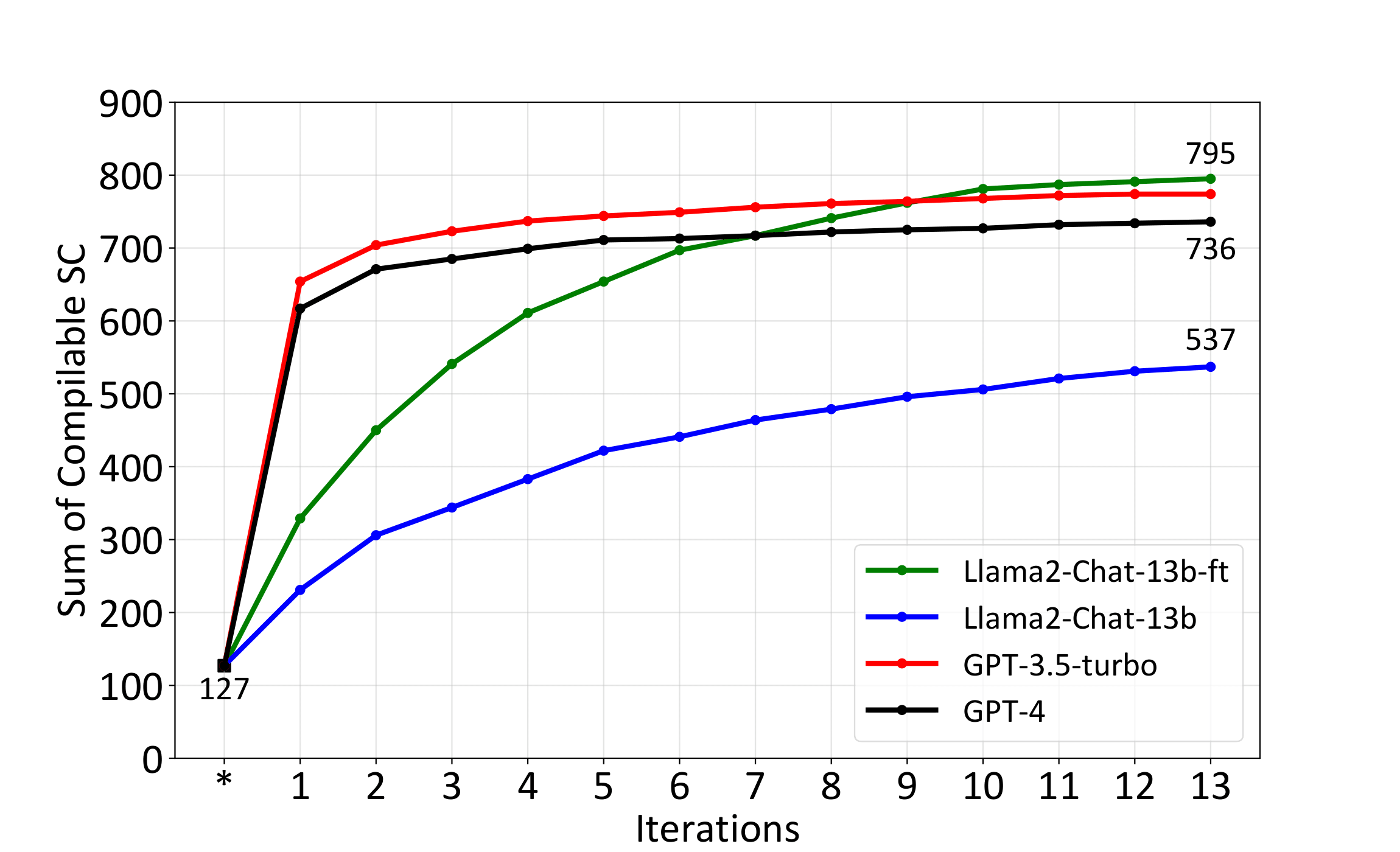}
    \caption{Performance of different models for code completion. The point with a horizontal coordinate of `*' on the graph represents the number of code that could have been directly compiled.}
    \label{fig:model_comparison}
\end{figure}

\noindent{\bf{Cost.}} Table ~\ref{tab:iterating} shows the costs of the 4 models on this task. Column `Time' shows the time required for each model to perform a complete round of code completion. It can be seen that the single round code completion time of \textit{GPT-4} reached 11.2 hours, nearly twice of our model and three times of the base model. Due to the need to pay a fee for each API call~\cite{apiprice}, the economic cost of GPT series models also counts. The total cost required for \textit{GPT-4} is the highest, reaching \$57.88, followed by \textit{GPT-3.5-turbo} at \$8.27. On the contrary, whether it is the open-source \textit{llama2} base model or the model we have fine-tuned, they are all deployed locally and do not require any economic cost. In order to obtain more completed smart contracts, we consider using a compromise solution, which is to perform ablations with the inference results of other models based on the inference results of our \textit{Code Completer}. Finally, we successfully obtained 846 correctly completed code snippets.

We observed the remaining 51 code snippets that cannot be completed by any of the models. There may be the following reasons: 1) inherent flaws within the snippets themselves, such as syntax errors, data type mismatches, and multiple constructors. 2) code complexity also plays a crucial role. This complexity can be attributed to two aspects: the length and clarity of the code. Longer code snippets are more challenging for the model to complete, as they require memorizing and understanding numerous details. Additionally, unclear code snippets, often containing undefined or unconventional variable and function names (e.g., \textit{foobar}, \textit{abcd}), further complicate the model’s ability to comprehend and complete the code. While many smart contract codes use traceable naming conventions that aid in logical inference (e.g., \textit{transfer}, \textit{withdraw}), some snippets from \textit{SO} employ non-standard names, making them harder for the model to interpret.


\begin{center}
    \begin{myboxc} \textbf{Answer to RQ1:} \textit{Code Completer} completed 75.5\% of the smart contract snippets correctly, outperforming both its base model and the GPT models, while also offering a lower usage cost compared to the GPT models.
    \end{myboxc}
\end{center}

\begin{table}[htbp]
    \centering
    \caption{Comparison of performance between \textit{Vulnerability Detector} and other tools (\textit{w.a.} F1 denotes weighted average F1 score for all vulnerabilities, and \# NUM denotes the number of contracts containing corresponding vulnerabilities).}
    \label{tab:tool_comparison}
    \resizebox{\columnwidth}{!}{
    \begin{tabular}{c|c|c c c c c c c}
        \hline
        \multicolumn{2}{c|}{\textbf{Vulnerability}} & RE & AC & AI & URV & DoS & BR & TM \\
        \hline
        \rowcolor{gray!25}
        \multicolumn{2}{c|}{\textbf{\# Num}} & 2 & 6 & 10 & 15 & 2 & 8 & 22 \\
        \hline
        \multicolumn{1}{c|}{\multirow{3}{*}{\textit{Conkas}}} 
        & \# TP & 0 & - & 0 & 0 & - & - & 0 \\
        \multicolumn{1}{c|}{} & \# FP & 22 & - & 17 & 0 & - & - & 1 \\
        \multicolumn{1}{c|}{} & \# FN & 2 & - & 10 & 15 & - & - & 22 \\
        \rowcolor{gray!15}
        \multicolumn{1}{c|}{\textit{w.a.} F1: 0} & F1 \% & 0 & - & 0 & 0 & - & - & 0 \\
        \hline
        \multicolumn{1}{c|}{\multirow{3}{*}{\textit{Maian}}} 
        & \# TP & - & 1 & - & - & - & - & - \\
        \multicolumn{1}{c|}{} & \# FP & - & 2 & - & - & - & - & - \\
        \multicolumn{1}{c|}{} & \# FN & - & 5 & - & - & - & - & - \\
        \rowcolor{gray!15}
        \multicolumn{1}{c|}{\textit{w.a.} F1: 22.2\%} & F1 \% & - & 22.2 & - & - & - & - & - \\
        \hline
        \multicolumn{1}{c|}{\multirow{3}{*}{\textit{Mythril}}} 
        & \# TP & 0 & 2 & 0 & 1 & - & - & - \\
        \multicolumn{1}{c|}{} & \# FP & 6 & 3 & 3 & 0 & - & - & - \\
        \multicolumn{1}{c|}{} & \# FN & 2 & 4 & 10 & 14 & - & - & - \\
        \rowcolor{gray!15}
        \multicolumn{1}{c|}{\textit{w.a.} F1: 12.3\%} & F1 \% & 0 & 36.4 & 0 & 12.5 & - & - & - \\
        \hline
        \multicolumn{1}{c|}{\multirow{3}{*}{\textit{Osiris}}} 
        & \# TP & 0 & - & 0 & - & 0 & - & 0 \\
        \multicolumn{1}{c|}{} & \# FP & 0 & - & 10 & - & 0 & - & 0 \\
        \multicolumn{1}{c|}{} & \# FN & 2 & - & 10 & - & 2 & - & 22 \\
        \rowcolor{gray!15}
        \multicolumn{1}{c|}{\textit{w.a.} F1: 0} & F1 \% & 0 & - & 0 & - & 0 & - & 0 \\
        \hline
        \multicolumn{1}{c|}{\multirow{3}{*}{\textit{Oyente}}} 
        & \# TP & 0 & 0 & 0 & - & 0 & - & 0 \\
        \multicolumn{1}{c|}{} & \# FP & 0 & 0 & 26 & - & 0 & - & 0 \\
        \multicolumn{1}{c|}{} & \# FN & 2 & 6 & 10 & - & 2 & - & 22 \\
        \rowcolor{gray!15}
        \multicolumn{1}{c|}{\textit{w.a.} F1: 0} & F1 \% & 0 & 0 & 0 & - & 0 & - & 0 \\
        \hline
        \multicolumn{1}{c|}{\multirow{3}{*}{\textit{Securify}}} 
        & \# TP & 0 & 0 & - & 0 & - & - & - \\
        \multicolumn{1}{c|}{} & \# FP & 0 & 1 & - & 0 & - & - & - \\
        \multicolumn{1}{c|}{} & \# FN & 2 & 6 & - & 15 & - & - & - \\
        \rowcolor{gray!15}
        \multicolumn{1}{c|}{\textit{w.a.} F1: 0} & F1 \% & 0 & 0 & - & 0 & - & - & - \\
        \hline
        \multicolumn{1}{c|}{\multirow{3}{*}{\textit{Slither}}}
        & \# TP & 0 & 1 & - & 0 & 0 & - & 1 \\
        \multicolumn{1}{c|}{} & \# FP & 1 & 3 & - & 10 & 1 & - & 0 \\
        \multicolumn{1}{c|}{} & \# FN & 2 & 5 & - & 15 & 2 & - & 21 \\
        \rowcolor{gray!15}
        \multicolumn{1}{c|}{\textit{w.a.} F1: 6.6\%} & F1 \% & 0 & 20.0 & - & 0 & 0 & - & 8.7 \\
        \hline
        \multicolumn{1}{c|}{\multirow{3}{*}{\textit{Smartcheck}}} 
        & \# TP & 0 & 0 & 0 & 3 & 1 & - & 0 \\
        \multicolumn{1}{c|}{} & \# FP & 0 & 0 & 3 & 5 & 29 & - & 0 \\
        \multicolumn{1}{c|}{} & \# FN & 2 & 6 & 10 & 12 & 1 & - & 22 \\
        \rowcolor{gray!15}
        \multicolumn{1}{c|}{\textit{w.a.} F1: 7.1\%} & F1 \% & 0 & 0 & 0 & 26.1 & 6.3 & - & 0 \\
        \hline
        \multicolumn{1}{c|}{\multirow{3}{*}{\textsc{SOChecker}}} 
        & \# TP & 1 & 4 & 8 & 9 & 1 & 5 & 20 \\
        \multicolumn{1}{c|}{} & \# FP & 0 & 0 & 1 & 0 & 0 & 0 & 0 \\
        \multicolumn{1}{c|}{} & \# FN & 1 & 2 & 2 & 6 & 1 & 3 & 2 \\
        \rowcolor{gray!15}
        \multicolumn{1}{c|}{\textit{w.a.} F1: \textbf{83.4\%}} & F1 \% & \textbf{66.7} & \textbf{80.0} & \textbf{84.2} & \textbf{75.0} & \textbf{66.7} & \textbf{76.9} & \textbf{95.2} \\
        \hline
    \end{tabular}
    }
\end{table}

\subsection{RQ2: Effectiveness of \textit{Vulnerability Detector}}
\label{subsec:effectiveness}
We assess the efficacy of \textit{Vulnerability Detector} using 846 complete code snippets, as referenced in Section~\ref{completion}, which are all correctly completed by the models. Given that nearly all other SOTA tools for vulnerability detection are designed for complete, compiled smart contract code, we can facilitate a fair comparison between our vulnerability detectors and these tools using this subset. 

We employed SmartBugs~\cite{diAngeloEtAl2023ASE}, a comprehensive framework that consolidates various smart contract vulnerability detection tools, to execute the evaluations. The selection criteria for these tools were: (1) They target Solidity source code. (2) 
\textit{Smartbugs} establishes a clear mapping rule between the vulnerability naming of the tool and DASP10 vulnerability naming (Because different tools may employ different naming for the same vulnerability). (3) The tool detects at least one vulnerability listed in the DASP10~\cite{dasp2018}. Consequently, we selected the following tools for our study: \textit{Conkas}~\cite{conkas}, \textit{Maian}~\cite{maian}, \textit{Mythril}~\cite{consensysmythril}, \textit{Osiris}~\cite{osiris}, \textit{Oyente}~\cite{oyente}, \textit{Securify}~\cite{securify}, \textit{Slither}~\cite{slither}, \textit{Honeybadger}~\cite{honeybadger} and \textit{Manticore}~\cite{mossberg2019manticore}. 

Our experiments were carried out on a Windows 11 system equipped with a 12th generation Intel i7 processor, 16GB of RAM, and a 5-minute timeout limit~\cite{chen2023chatgpt} for each tool. Two experienced smart contract researchers independently annotated the presence of vulnerabilities in all snippets of the smart contract. In instances of disagreement, two researchers engaged in discussions to reconcile and unify their conclusions. We used seven metrics to evaluate the experimental results, namely true positive (TP), true negative (TN), false positive (FP), false negative (FN), precision, recall and F1 score. TP and TN represent smart contracts with certain vulnerabilities correctly detected by our \textit{Vulnerability Detector} and smart contracts without certain vulnerabilities, respectively. FP and FN indicate that \textit{Vulnerability Detector} has incorrectly detected smart contracts with or without certain vulnerabilities. We calculate precision using formula $P=TP/(TP+FP)$, recall using formula $R=TP/(TP+FN)$, F1 using formula $F1=2*P*R/(P+R)$. For w.a. F1, we calculate it as follows: $w.a. F1=\frac{\sum\nolimits_{i=1}^{n}v_i*F1_i}{\sum\nolimits_{i=1}^{n}v_i}$, where $n$ represents the type of vulnerability and $v_i$ represents the number of vulnerabilities.

Table~\ref{tab:tool_comparison} presents the results of our experiments. Tools \textit{Honeybadger} and \textit{Manticore} are absent from Table~\ref{tab:tool_comparison} as they failed to identify any positive cases within the dataset. Due to the absence of \textit{Front Running} and \textit{Short Address Attack} vulnerabilities in our dataset, we also omitted them in Table~\ref{tab:tool_comparison}. Our experimental results reveal that our \textit{Vulnerability Detector} obtains the highest weighted average F1 score of 83.4\%, surpassing other SOTA tools across all indicators. A comparative analysis with other tools reveals that our \textit{Vulnerability Detector} not only encompasses all DASP10~\cite{dasp2018} vulnerabilities but also exhibits superior detection performance across all categories, underscoring its high efficacy. However, \textit{Vulnerability Detector} still generate some false alarms. We examined each of them individually and found that they were caused by several factors. \textit{Z3} has inherent difficulties in handling complex path conditions, particularly those involving factorial or loop operations, which can easily lead to timeouts and affect the acquisition of critical path information. Additionally, different compiler versions can influence detection results. For instance, \textit{Solidity} v0.8.x includes default integer overflow checks, which differ from earlier versions. \textit{Vulnerability Detector} relies on predefined vulnerability patterns and may not account for these optimizations, leading to discrepancies with actual vulnerabilities.
\begin{center}
    \begin{myboxc} \textbf{Answer to RQ2:} Our \textit{Vulnerability Detector} achieved an average F1 score of 83.4\% on the dataset. Compared to the 10 state-of-the-art tools, it not only identifies the most types of vulnerability listed in DASP10, but also demonstrates the best detection performance for each type of vulnerability.
    \end{myboxc}
\end{center}

\subsection{RQ3: Effectiveness of \textsc{SOChecker}}
We evaluated   \textsc{SOChecker}’s overall performance with authentic code snippets sourced from \textit{SO}. We executed \textsc{SOChecker} on 897 code snippets collected from \textit{SO}; all experimental settings, such as 
temperature, maintain the same as them in RQ1 and RQ2.

\begin{figure}[htbp]
\setlength{\abovecaptionskip}{0.2cm}
\setlength{\belowcaptionskip}{-0.1cm}
    \centering
    \includegraphics[width=\columnwidth]{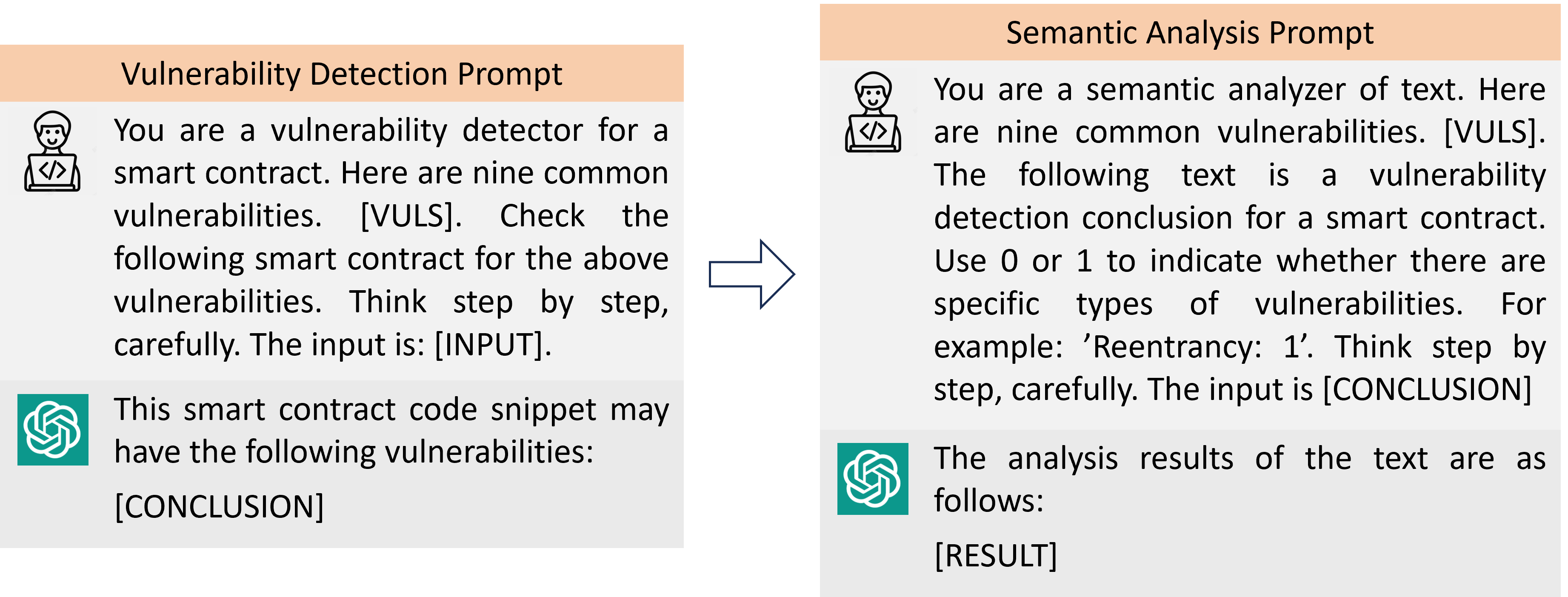}
    \caption{Example of using LLM to detect vulnerabilities.}
    \label{fig:prompt}
\end{figure}

LLMs can also be directly used for vulnerability detection of smart contract snippets. Consequently, we employed GPT-3.5 and GPT-4 to analyse the same dataset, facilitating a comparative analysis of their performance against \textsc{SOChecker}. Specifically, we designed a prompt informed by others' previous work~\cite{chen2023chatgpt} to obtain non-binary results (i.e., in LLM's response, the presence of vulnerabilities is indicated by "1", while their absence is denoted by "0".). We then gave these results back to \textit{GPT-4} for semantic analysis, so that binary results about the existence of these vulnerabilities can be obtained. In Figure~\ref{fig:prompt}, within the vulnerability detection prompt, ``[VULS]'' denotes the names of all vulnerabilities, ``[Input]'' specifies the target code subject to detection, and "[CONCLUSION]" represents the detection outcome provided by ChatGPT. For the semantic analysis prompt, ``[VULS]'' continues to signify the vulnerability name, ``[CONCLUSION]'' refers to the conclusion derived from prior vulnerability detection, and ``[RESULT]'' indicates the semantic analysis result delivered by ChatGPT.

Table~\ref{tab:gpt_comparison} displays the performance of \textsc{SOChecker} and GPT series models in detecting vulnerabilities across 897 real smart contract snippets. The weighted average F1 score achieved by \textsc{SOChecker} is 68.2\%, in contrast to GPT-3.5 and GPT-4, which scored 20.9\% and 33.2\%, respectively. Although GPTs achieving high recall rates for most vulnerabilities, their precision remains low, leading to suboptimal overall performance. This finding aligns with the outcomes of prior research~\cite{chen2023chatgpt}. We observed that \textsc{SOChecker}'s recall rate was relatively low. Upon individually analyzing each false negative, we discovered that the majority were attributed to the model either patching the original vulnerabilities or altering the original semantics during code completion. Such issues prove challenging for pruning algorithms to effectively address.

\begin{table}[hbtp]
\setlength{\abovecaptionskip}{0cm}
\setlength{\belowcaptionskip}{0cm}
    \centering

    \caption{Comparison of performance between \textsc{SOChecker} and GPTs.}
    \label{tab:gpt_comparison}
    \resizebox{\linewidth}{!}{
    \begin{tabular}{c|c|c|c|c|c|c|c|c|c}
        \hline
        \multirow{2}{*}{\textbf{Vuls}} & \multicolumn{3}{c|}{\cellcolor{gray!30}\textit{GPT-3.5}} & \multicolumn{3}{c|}{\cellcolor{gray!30}\textit{GPT-4}} & \multicolumn{3}{c}{\cellcolor{gray!30}\textsc{SOChecker}} \\
        \hhline{~---------}
        & \multicolumn{1}{c}{\cellcolor{gray!15}$P\%$} & \multicolumn{1}{c}{\cellcolor{gray!15}$R\%$} & \multicolumn{1}{c|}{\cellcolor{gray!15}$F1\%$} & \multicolumn{1}{c}{\cellcolor{gray!15}$P\%$} & \multicolumn{1}{c}{\cellcolor{gray!15}$R\%$} & \multicolumn{1}{c|}{\cellcolor{gray!15}$F1\%$} & \multicolumn{1}{c}{\cellcolor{gray!15}$P\%$} & \multicolumn{1}{c}{\cellcolor{gray!15}$R\%$} & \multicolumn{1}{c}{\cellcolor{gray!15}$F1\%$} \\
        \hline
        RE & \multicolumn{1}{c}{1.8} & \multicolumn{1}{c}{100} & \multicolumn{1}{c|}{3.5} & \multicolumn{1}{c}{2.0} & \multicolumn{1}{c}{100} & \multicolumn{1}{c|}{4.0} & \multicolumn{1}{c}{100} & \multicolumn{1}{c}{50.0} & \multicolumn{1}{c}{\textbf{66.7}} \\
        AC & \multicolumn{1}{c}{0.8} & \multicolumn{1}{c}{50.0} & \multicolumn{1}{c|}{1.7} & \multicolumn{1}{c}{1.6} & \multicolumn{1}{c}{83.3} & \multicolumn{1}{c|}{3.2} & \multicolumn{1}{c}{100} & \multicolumn{1}{c}{66.7} & \multicolumn{1}{c}{\textbf{80.0}} \\
        AI & \multicolumn{1}{c}{7.3} & \multicolumn{1}{c}{33.3} & \multicolumn{1}{c|}{12.0} & \multicolumn{1}{c}{8.1} & \multicolumn{1}{c}{50.0} & \multicolumn{1}{c|}{14.0} & \multicolumn{1}{c}{88.9} & \multicolumn{1}{c}{44.4} & \multicolumn{1}{c}{\textbf{59.3}} \\
        URV & \multicolumn{1}{c}{11.4} & \multicolumn{1}{c}{65.0} & \multicolumn{1}{c|}{19.4} & \multicolumn{1}{c}{26.3} & \multicolumn{1}{c}{75.0} & \multicolumn{1}{c|}{39.0} & \multicolumn{1}{c}{100} & \multicolumn{1}{c}{45.0} & \multicolumn{1}{c}{\textbf{62.1}} \\
        DoS & \multicolumn{1}{c}{1.6} & \multicolumn{1}{c}{50.0} & \multicolumn{1}{c|}{3.0} & \multicolumn{1}{c}{1.9} & \multicolumn{1}{c}{100} & \multicolumn{1}{c|}{3.7} & \multicolumn{1}{c}{50.0} & \multicolumn{1}{c}{25.0} & \multicolumn{1}{c}{\textbf{33.3}} \\
        BR & \multicolumn{1}{c}{17.9} & \multicolumn{1}{c}{70.0} & \multicolumn{1}{c|}{28.6} & \multicolumn{1}{c}{30.8} & \multicolumn{1}{c}{80.0} & \multicolumn{1}{c|}{44.4} & \multicolumn{1}{c}{83.3} & \multicolumn{1}{c}{50.0} & \multicolumn{1}{c}{\textbf{62.5}} \\
        TM & \multicolumn{1}{c}{22.7} & \multicolumn{1}{c}{63.0} & \multicolumn{1}{c|}{33.3} & \multicolumn{1}{c}{37.5} & \multicolumn{1}{c}{77.8} & \multicolumn{1}{c|}{50.6} & \multicolumn{1}{c}{95.2} & \multicolumn{1}{c}{74.1} & \multicolumn{1}{c}{\textbf{83.3}} \\
        \hline
        \rowcolor{gray!20}
        \textit{w.a.} & \multicolumn{1}{c}{13.4} & \multicolumn{1}{c}{57.5} & \multicolumn{1}{c|}{20.9} & \multicolumn{1}{c}{23.1} & \multicolumn{1}{c}{73.6} & \multicolumn{1}{c|}{33.2} & \multicolumn{1}{c}{92.0} & \multicolumn{1}{c}{55.2} & \multicolumn{1}{c}{\textbf{68.2}} \\
        \hline
    \end{tabular}
    }
\end{table}

We tried to execute other vulnerability detection tools discussed in Section~\ref{subsec:effectiveness} on this code snippets dataset for comparative analysis. However, as these traditional tools are designed to analyze complete smart contracts, they encountered errors with 770 code snippets presented in fragmentary form. Of the remaining 127 complete contracts, encompassing a total of 18 vulnerabilities, only \textit{Slither} and \textit{Smartcheck} managed to detect 1 and 2 TPs, respectively, while \textsc{SOChecker} identified 14. This outcome suggests that, in the context of fragmented code, our tool demonstrates greater utility compared to conventional tools.

\begin{center}
    \begin{myboxc} \textbf{Answer to RQ3:} SOChecker achieved an average F1 score of 68.2\% on real datasets sourced from \textit{SO}, surpassing both GPT models and other traditional vulnerability detection tools.
    \end{myboxc}
\end{center}


\section{Threats to Validity}
\noindent {\bf{Internal Threats.}} A potential internal threat in our study is the reliance on specific tags for collecting posts from \textit{SO}. This method may overlook relevant discussions that do not feature the designated tags. However, by selecting widely-used tags such as ``Solidity'', ``Ethereum'', ``ERC20'', ``ERC721'' and ``Contract'', we aim to capture a diverse range of smart contract-related content. The extensive volume of posts gathered from these tags helps mitigate this risk and ensures comprehensive coverage for analysis.

\noindent {\bf{External Threats.}} We designed a survey to assess whether developers had implemented risky \textit{SO} code into their projects. However, we did not directly trace these codes on the Ethereum chain due to the vast amount of contract information available, which made it impractical to complete within our timeframe. 
Nonetheless, the data gathered from the survey can provide valuable insights into the real-world scenario. Our survey targets smart contract practitioners with diverse backgrounds and levels of experience, allowing them to provide feedback on their use of \textit{SO} code.
\section{Related Work}
\label{sec:rw}
Due to the rise of large language models, some scholars have recently conducted in-depth research on them. Fan et al.~\cite{fan2023automated} studied whether automatic program repair technology can fix the error solutions generated by LLMs in the \textit{LeetCode} competition. Li et al.~\cite{li2023test} studied the limitations of LLMs in generating software failure-induced test cases and proposed a differential prompt method to improve effectiveness. 
Ma et al.~\cite{ma2023scope} evaluated the performance of \textit{ChatGPT} in various subdomains of software engineering. Chen et al.~\cite{chen2023chatgpt} evaluated the performance of \textit{ChatGPT} in detecting vulnerabilities in smart contracts. David et al.~\cite{david2023need} studied the detection ability of LLMs such as \textit{Claude} and \textit{GPT} for actual attacks on smart contracts.


Smart contract vulnerability detection has always been a research hotspot in the fields of blockchain and smart contract security. Many work uses static analysis methods to detect potential vulnerabilities in code before contract deployment~\cite{oyente,smartcheck,securify,slither,confuzzius,sailfish}. Some work has also pointed out the problems with current mainstream tools~\cite{zheng2023turn}. Fuzzing is also a commonly used method to detect vulnerabilities in smart contracts~\cite{grieco2020echidna,jiang2018contractfuzzer,wustholz2020harvey}
In addition, machine learning has also been used in smart contract vulnerability detection tasks in recent years~\cite{liao2019soliaudit,sendner2023smarter}. 
Our approach to vulnerability detection aligns with established methodologies but offers unique features compared to previous work~\cite{oyente,smartcheck,securify,slither,confuzzius,sailfish}. Firstly, it possesses the capability to prune programs effectively, rendering it adept at handling the fragmented code commonly found on \textit{SO}. Secondly, it simulates program execution with greater completeness, capturing a wider array of operation codes. Thirdly, given that the majority of the targets are simple contracts, efficiency optimization is not a primary concern, allowing us to incorporate detailed mechanisms like memory and storage mapping.

The code issues on the famous developer forum \textit{Stack Overflow (SO)} have received increasing attention from researchers in recent years. Despite previous efforts~\cite{zhang2021study,verdi2020empirical,meldrum2020understanding} to perform security analysis on \textit{SO} code, most of them focus on traditional programming languages (e.g., \textit{C/C++}, \textit{Java}), while \textit{Solidity}, the most popular programming language for smart contracts~\cite{dannen2017introducing}, has received less attention. In addition, these works are mainly completed by manual code inspection, leaving a gap in automated vulnerability detection for Solidity-based smart contracts. Zhang et al.~\cite{zhang2021study} empirically studied the prevalence of the Common Weakness Enumeration (CWE), in code snippets of \textit{C/C++} related answers. Verdi et al.~\cite{verdi2020empirical} investigated security vulnerabilities in \textit{C++} code snippets on \textit{SO} over a period of 10 years. Meldrum et al.~\cite{meldrum2020understanding} evaluated the quality of \textit{SO} code in various aspects, including reliability and conformance to programming rules, readability, performance and security. 
\section{Conclusion}
\label{sec:conclusion}


We conducted a survey to investigate their usage patterns and perspectives regarding smart contract code snippets on \textit{Stack Overflow} and obtained feedback from 74 smart contract practitioners. Our findings suggest a significant risk associated with the adoption of vulnerable code snippets by developers, potentially compromising the security of the blockchain ecosystem. We wanted to support developers to identify such vulnerabilities within smart contract code snippets. To do this we introduced \textsc{SOChecker}, a novel tool that combines a fine-tuned \textit{Llama2}-based \textit{Code Completer} with a \textit{Vulnerability Detector}.  Tested on 897 code snippets, \textsc{SOChecker} demonstrated greatly superior performance over existing GPT-serires LLMs and other program analysis tools.

\section*{Acknowledgements}
This work is partially supported by fundings from the National Key R\&D Program of China (2022YFB2702203), the National Natural Science Foundation of China (62302534, 62332004).
\normalem





\bibliographystyle{ACM-Reference-Format}


\end{document}